\documentclass[12pt,eadjoint tfnpsf]{article}

\catcode`\@=11
\@addtoreset{equation}{section}

\global\arraycolsep=1pt

\usepackage{amsbsy,amssymb,latexsym,amsfonts}
\usepackage{mathrsfs}
\usepackage{graphicx}

\RequirePackage[dvips,usenames]{color}
\definecolor{fireblick}{rgb}{0.698039,0.133333,0.133333}

\newcommand{\beq}{\begin{equation}}
\newcommand{\eeq}{\end{equation}}
\newcommand{\bea}{\begin{eqnarray}}
\newcommand{\eea}{\end{eqnarray}}

\newcommand{\CD}{{\mathcal D}}
\newcommand{\CF}{{\mathcal F}}
\newcommand{\CL}{{\mathcal L}}
\newcommand{\CN}{{\mathcal N}}
\newcommand{\CO}{{\mathcal O}}

\newcommand{\CS}{{\mathcal S}}

\newcommand{\CW}{{\mathcal W}}

\newcommand\fD{\mathfrak D}

\def\Im{\mathop{\rm Im}}
\renewcommand\Re{{\mathrm{Re}}}

\newcommand\ui{{\underline{i}}}
\newcommand\uj{{\underline{j}}}

\def\Tr{\mathop{\rm Tr}}

\newcommand\diag{\mathrm{diag}}

\begin{document}
%
%
\begin{titlepage}
\begin{flushright}
\normalsize
~~~~
OCU-PHYS 320\\
YITP-09-58\\
September, 2009 \\
\end{flushright}

\vspace{40pt}

\begin{center}
{\LARGE Low Energy Processes Associated with} \\
\vspace{8pt}
{\LARGE Spontaneously Broken $\CN=2$ Supersymmetry}\\
\end{center}

\vspace{15pt}

\begin{center}
{%
H. Itoyama$^{a, b}$\footnote{e-mail: itoyama@sci.osaka-cu.ac.jp},
K. Maruyoshi$^c$\footnote{e-mail: maruyosh@yukawa.kyoto-u.ac.jp}
and S. Minato$^{a}$\footnote{e-mail: minato@sci.osaka-cu.ac.jp}
}\\
%
\vspace{18pt}
%
$^a$ \it Department of Mathematics and Physics, Graduate School of Science\\
Osaka City University\\
\vspace{5pt}

$^b$ \it Osaka City University Advanced Mathematical Institute (OCAMI)

\vspace{5pt}

3-3-138, Sugimoto, Sumiyoshi-ku, Osaka, 558-8585, Japan \\

\vspace{10pt}

$^c$ \it Yukawa Institute for Theoretical Physics, Kyoto University, Kyoto 606-8502, Japan\\
\end{center}
%
\vspace{15pt}
%
\begin{center}
Abstract\\
\end{center}
  We consider low energy processes described by the $\CN=2$ supercurrent on its partially (to $\CN =1$)
  and spontaneously broken vacuum and the attendant Nambu-Goldstone fermion (NGF), 
  which the presence of the electric and magnetic Fayet-Iliopoulos (FI) terms is responsible for.
  We show suppressions of amplitudes decaying into the NGF as its momentum becomes small.
  In the lagrangian realization (namely, the model of arXiv:hep-th/0409060) of the conserved supercurrent, 
  the NGF resides in the overall $U(1)$, which is nonetheless not decoupled, and interacts with the $SU(N)$ sector
  through nonderivative as well as derivative couplings.  
  The low energy suppression is instead accomplished by a cancellation 
  between the annihilation diagram from the Yukawa couplings and the contact four-Fermi terms.
  We give a complete form of the supercurrent and the model is recast in more transparent notation. 

\vfill

\setcounter{footnote}{0}
\renewcommand{\thefootnote}{\arabic{footnote}}

\end{titlepage}

\section{Introduction}
\label{sec:intro}
  Notion of spontaneously broken symmetry \cite{Nambu, NJL} 
  and the current algebras of sixties \cite{AdlerDashen} had successes 
  in investigating some of the low energy properties of hadrons. 
  A basic argument starts from stating that 
  the diagrams representing the hadronic matrix elements of the conserved current fall into two categories - 
  the one in which the current couples a nearly on-shell massless Nambu-Goldstone pion 
  that propagates before it interacts with initial and final hadrons 
  and the other in which the current couples directly to the hadrons. 
  There is no candidate excitation to produce a singularity in the latter category.
  The conservation law then forces the residue of the pion pole at the first category to vanish at zero momentum, 
  which in turn implies that emission amplitudes of the pion vanish at zero momentum.
  The same logic can be applied to the case of spontaneously broken $\CN=1$ supersymmetry 
  and emission amplitudes of the Nambu-Goldstone fermion (NGF). 
  As was shown by de Wit-Freedman and Bardeen \cite{dWF, dWF2, Bardeen},
  this argument and the observed $\beta$ decay spectrum precluded the idea of the NGF being neutrino.
  (See also \cite{KK} for recent work on the Nambu-Goldstone boson low energy theorems in $\CN=8$ supergravity.)

  Turning our attention to extended supersymmetry,
  it is in fact straightforward to generalize the above argument to the case
  in which $\CN=2$ supersymmetry is partially and spontaneously broken to $\CN=1$.
  It has been known for a long time that partial breaking of $\CN=2$ supersymmetry to $\CN=1$ is accomplished 
  by the simultaneous presence of the electric and magnetic Fayet-Iliopoulos (FI) terms \cite{APT}.
  The lagrangian realization with a nonabelian $U(N)$ gauge group has been given for some time \cite{FIS1, FIS2, FIS3}. 
  See \cite{KP} for related topics.
  Changing the strength of the FI terms, we are able to interpolate \cite{IM, IM2, Fujiwara} 
  between $\CN=2$ super-Yang Mills and $\CN=1$ super-Yang Mills 
  with a superpotential consisting of a chiral multiplet in the adjoint representation.
  The corresponding interpolation of the low energy effective action (LEEA) offers
  an interesting arena for the study of the exact determination of the LEEA \cite{SW, Vafa, CIV, DV, DGLVZ, CDSW} 
  and the integrable systems \cite{Chekhov1}.
  Several pieces of work along this direction have already appeared 
  \cite{IM, Ferrari, IM2, ABSV, HMPS, Maruyoshi, FerrariWens, Maruyoshiextended}.
  (For closely related works, see \cite{LMN, MN, Marshakov, NNT}.)
  Matter hypermultiplets in the bi-fundamental representation have been included in the work of \cite{IMS}.
  Albeit being nonchiral, these, combined with several well-known mechanisms that break $\CN=1$ supersymmetry, 
  permit semi-realistic considerations beyond the standard model \cite{3books}.
  In fact, there are already some phenomenological works on $\CN=2$ supersymmetric models 
  which are relevant in ten TEV energy scale \cite{ChoiDrees}:
  partial breaking of $\CN=2$ supersymmetry provides an interesting prospect to physics issues at LHC.
  
  In models with rigid $\CN=2$ supersymmetry realized on the partially broken vacua, a massless NGF is predicted 
  while, in local models, the NGF is absorbed into a massive gravitino by the super-Higgs mechanism. 
  In $\CN=2$ supersymmetry, the rigid models and the local models live
  in different Kahler geometries (special Kahler v.s. quaternionic Kahler) \cite{Andrianopoli}.
  For prescription of partial breaking of local $\CN=2$ supersymmetry, see \cite{sugra}.
  In general, models with the FI terms are known to be hard to couple to gravity.
  For a recent discussion, see \cite{KS}.
  In this paper, we have in mind the rigid realization of partially broken $\CN=2$ supersymmetry.
  
  There is apparently, however, a puzzling situation which we encounter with regard to some properties of the model 
  and suppression of zero momentum emission amplitudes of the NGF derived 
  from the conserved supercurrent on the broken vacuum.
  The overall $U(1)$ sector, which the NGF and its superpartner (dark photon) resides in 
  and which drives the breaking of $\CN=2$ to $\CN=1$, is successfully coupled to the $SU(N)$ sector.  
  The mass spectrum of the model consists of three types of $\CN=1$ supermultiplet belonging to the gauge group $U(N)$
  broken by the Higgs mechanism to a variety of product gauge groups \cite{FIS2}.
  We show explicitly in this paper the presence of non-derivative Yukawa couplings that involve the NGF. 
  It is an interesting question to address how this counterintuitive structure is compatible with the consequences 
  derived from the low energy theorem. 
  Answering this demystifies the situation.
  
  In the next section, we take the point of view of the algebra of currents 
  to consider matrix elements of the $\CN=2$ supercurrent that couples to the NGF.  
  The low energy theorem is given that  attendant multiparticle amplitudes emitting the NGF 
  (by the factorization of the pole) are suppressed as its momentum becomes small
  under the assumption that there is no other source of singularity in the limit.
  Actually a major source of the potential singularities is the mass degeneracy of the spectrum
  and one may be afraid in principle that unbroken $\CN=1$ supermultiplets may realize this possibility.
  In order to examine the validity of the assumption, we derive an explicit form of the supercurrent 
  from the model we discuss in the subsequent section. 
  We are able to argue that the singularity due to the degeneracy does not appear 
  by the insertion of the component of the $\CN =2$ supercurrent that couples to the NGF.
  In section \ref{sec:lagrangian}, we recall a lagrangian realization of the current algebra, 
  which is the model of \cite{FIS1} already mentioned.
  After recalling several properties  given in \cite{FIS2, Maruyoshi}, we derive interaction vertices of the NGF. 
  Among other things, we point out the presence of NGF-gaugino-scalar Yukawa couplings
  which do not disappear even at zero momentum transfer to the NGF.
  In section \ref{sec:LEsuppression}, 
  we consider an emission amplitude of the NGF directly from tree diagrams of the model, taking the simplest case.  
  We show that the suppression at zero momentum is realized by the cancellation 
  between the $s$-channel annihilation diagram and the four-fermi terms. 
  
  In Appendix \ref{sec:component}, we give the component lagrangian of the model in the new notation
  as compared with \cite{FIS1}.
  In Appendix \ref{sec:susytr}, we give the transformation law of the extended supersymmetry.
  In Appendix \ref{sec:N=1}, we review the low energy theorem associated with spontaneously broken $\CN =1$ supersymmetry.
  We adopt the notation of \cite{Lykken}.

\section{Low energy theorem associated with conserved $\CN=2$ supercurrent}
\label{sec:LEtheorem}

\subsection{Low energy suppression of processes with NGF emission}
\label{subsec:}
  Let us consider matrix elements of the $\CN=2$ supercurrent $({\cal S}^{(1)\mu}, {\cal S}^{(2)\mu})$. 
  As will be explained in subsection \ref{subsec:FISclassicalvacua}, 
  we focus on the vacua where the second $\CN=1$ supersymmetry
  corresponding to ${\cal S}^{(2)\mu}$ is broken for simplicity.
  In this choice, the NGF is coupled to ${\cal S}^{(2)\mu}$.
  It is straightforward to apply the analysis in the subsequent sections to the theory on the vacua
  where the first $\CN=1$ supersymmetry is broken.
  
  We consider the Fourier transform (F.T.) of the matrix element of ${\cal S}^{(2)\mu}$ is
    \bea
    {\rm F.T.} \langle p_{f} ; \cdots | {\cal S}^{(2)\mu} | p_{i} ; \cdots \rangle (q), \;\; q = p_i - p_f\;.
    \eea
  We have here adopted the majorana notation for the supercurrent:
    \bea
    {\cal S}^{(2)\mu} 
     =     ( {\cal S}^{(2)\mu}_{\alpha}, \bar{ {\cal S}}^{\dot{\alpha} (2)\mu})^{t} \;\;.
    \eea
  The explicit form of the supercurrent will be given below.
  For definiteness, the initial state is taken to be a multiparticle bosonic state with a set of momenta  $p_{i}$  
  while the final state to be a fermionic one with $p_{f}$.
  We have suppressed the spinor indices.
  
  The decomposition of this quantity after considering the on-shell condition should go as
    \bea 
    {\rm F.T.} \langle p_{f} ; \cdots | {\cal S}^{(2)\mu} | p_{i} ; \cdots \rangle (q)
     =     q^{\mu} F(q^2, \cdots) + R^{\mu}(q^2, \cdots)\;.
    \eea 
  In the special case where the initial state is a scalar and the final state is a spinor, this reads
    \bea
    {\rm F.T.} \langle p_{f} ; \cdots | {\cal S}^{(2)\mu} | p_{i} ; \cdots \rangle (q)
     =     q^{\mu} A_{1}(q^2, \cdots) \left( \bar{U}_{f}(p_{f}) C \right)^t
         + A_{2}(q^2, \cdots) \left( \bar{U}_{f}(p_{f}) \gamma^{\mu} C \right)^t \;.
    \eea
  Here the spinor $\bar{U}_{f}(p_{f})$ are the final state wave functions. 
  Imposing current conservation and noting that there is no singularity contributing to
  $R^{\mu}$ or $A_2$ in the limit $q \rightarrow 0$, we obtain 
    \bea
    \lim_{q^{\mu} \rightarrow 0 } q^2 F(q^2, \cdots)
     =     0 \;.
    \eea
  The residue of $\frac{1}{q^2}$ in $F$ or $A_1$ is $\langle f \lambda_{{\rm NGF}} | i \rangle$
  up to the numerical factor.
  The emission amplitude of the NGF is suppressed as  $q^{\mu}$ vanishes.
  
  The above simple argument is based on the assumption that there is no other singularity in this limit 
  and its validity needs to be examined. 
  In fact, in the Fayet model of broken $\CN=1$ supersymmetry \cite{FI}, 
  there exists a two-point coupling of photon and the NGF introduced by the insertion of the supercurrent \cite{dWF}.   
  The simultaneous emission amplitude of the NGF and photon is not suppressed as $q^{\mu} \rightarrow 0$. 
  More generally, the emission amplitude at zero momentum transfer will not be suppressed 
  if there is a term consisting of two fields whose masses are degenerate. 
  These are briefly illustrated in appendix \ref{sec:N=1}.
  
\subsection{$\CN=2$ supercurrent and its matrix elements}
\label{subsec:LEsupercurrent}
  In order to settle down the issue raised in the last subsection, 
  it is preferable to have an explicit form of the supercurrent. 
  Let us exploit the one derived from the lagrangian realization discussed in later sections. 
  
  The Noether currents associated with the first $\CN=1$ supersymmetry and the second one are respectively
    \bea
    \eta_1 \CS^{(1) \mu}
    &=&    \sqrt{2} g_{ab} \eta_1 \sigma^\nu \bar{\sigma}^\mu \psi^a \CD_\nu \bar{\phi}^b
         + \frac{i}{2} g_{ab} \eta_1 \sigma^\rho \bar{\sigma}^\nu \sigma^\mu \bar{\lambda}^a F_{\nu \rho}^b
         - 2 i \sqrt{N} (\bar{e} \delta^0_a + m \bar{\CF}_{0a} ) \eta_1 \sigma^\mu \bar{\psi}^a
           \nonumber \\
    & &  - \frac{i}{2} g_{ab} f^b_{cd} \eta_1 \sigma^\mu \bar{\lambda}^a \bar{\phi}^c \phi^d
         - \frac{1}{2 \sqrt{2}} \CF_{abc} \eta_1 \psi^a (\psi^b \sigma^\mu \bar{\psi}^c)
         - \frac{1}{2 \sqrt{2}} \bar{\CF}_{abc} \eta_1 \sigma^\mu \bar{\psi}^a (\lambda^b \bar{\lambda}^c)
    \eea
  and 
    \bea
    \eta_2 \CS^{(2) \mu}
    &=&  - \sqrt{2} g_{ab} \eta_2 \sigma^\nu \bar{\sigma}^\mu \lambda^a \CD_\nu \bar{\phi}^b
         + \frac{i}{2} g_{ab} \eta_2 \sigma^\rho \bar{\sigma}^\nu \sigma^\mu \bar{\psi}^a F_{\nu \rho}^b
         + 2 i \sqrt{N} (e \delta^0_a + m \bar{\CF}_{0a} ) \eta_2 \sigma^\mu \bar{\lambda}^a
           \nonumber \\
    & &  - \frac{i}{2} g_{ab} f^b_{cd} \eta_2 \sigma^\mu \bar{\psi}^a \bar{\phi}^c \phi^d
         + \frac{1}{2 \sqrt{2}} \CF_{abc} \eta_2 \lambda^a (\lambda^b \sigma^\mu \bar{\lambda}^c)
         + \frac{1}{2 \sqrt{2}} \bar{\CF}_{abc} \eta_2 \sigma^\mu \bar{\lambda}^a (\psi^b \bar{\psi}^c).
           \label{susycurrent2}
    \eea
  The terms in $\CS^{(2) \mu}$ relevant to our present discussion are
  $- \sqrt{2} g_{ab} \eta_2 \sigma^\nu \bar{\sigma}^\mu \lambda^a \partial_\nu \bar{\phi}^b$,
  $\frac{i}{2} g_{ab} \eta_2 \sigma^{ [ \rho} \bar{\sigma}^{\nu]} \sigma^\mu \bar{\psi}^a \partial_{\nu} A_{\rho}^b$
  and $2 i \sqrt{N} m \langle \bar{\CF}_{0ab} \rangle \bar{\phi}^b  \eta_2 \sigma^\mu \bar{\lambda}^a$.
  All of these terms contain one massless field and one massive field.
  None of them contains any mass degeneracy. The singularity is not introduced by the mass degeneracy.
  We have not thoroughly investigated the consequences derived from the last two terms of (\ref{susycurrent2})
  which crate three-point couplings.

\section{Lagrangian realization of partially and spontaneously broken $\CN=2$ supersymmetry}
\label{sec:lagrangian}

\subsection{The model and the transformation law}
\label{subsec:FIS}
  We first recast the lagrangian of the $U(N)$ gauge model with partially and spontaneously broken 
  $\CN=2$ supersymmetry of \cite{FIS1, FIS2} in more transparent notation:
    \bea
    \CL_{U(N)}
     =     \Im 
           \left[ 
           \int d^4 \theta 
           {\rm Tr} 
           \bar{\Phi} e^{ad V} 
           \frac{\partial \CF(\Phi)}{\partial \Phi}
         + \int d^2 \theta
           \frac{1}{2} 
           \frac{\partial^2 \CF(\Phi)}{\partial \Phi^a \partial \Phi^b}
           \CW^{\alpha a} \CW^b_{\alpha}
           \right]
         + \left(\int d^2 \theta W(\Phi)
         + c.c. \right),      
           \label{FIS'}
    \eea
  where the superpotential is
    \bea
    W(\Phi)
     =     \Tr \left( 2 e \Phi
         + m \frac{\partial \CF(\Phi)}{\partial \Phi} \right).
           \label{prepot}
    \eea
  In this notation, the electric Fayet-Iliopoulos parameter $e$ is complex while the magnetic one $m$ is real. 
  Both terms are vectors under $SU(2)_R$.
  An apparent difference from the original notation of \cite{APT, FIS3} just translates 
  into a different way of fixing this rigid $SU(2)_R$ rotation.
  In terms of $U(N)$ generators $t_a$, $a= 0, \ldots, N^2 -1$,
  ($a = 0$ refers to the overall $U(1)$ generator)\footnote{
    We normalize the generators as $\Tr (t_a t_b) = \delta_{ab}/2$.
    In this normalization, the $U(1)$ generator becomes $t_0 = \frac{1}{\sqrt{2 N}} 1_{N \times N}$.}, 
  the superfield $\Psi = \{ V, \Phi \}$ is expanded as $\Psi = \Psi^a t_a$.
  In what follows, we will denote the derivatives of $F(\Phi)$ 
  with respect to $\Phi^a, \Phi^b, \ldots$ by $F_{ab \ldots}$.
  The lagrangian in the component fields is explicitly written in appendix \ref{sec:component}.
  
  This lagrangian is invariant under $\CN=2$ supersymmetry.
  Invariance under $\CN=1$ supersymmetry is manifest: $\delta^{(1)} \CL = 0$.
  Another $\CN=1$ supersymmetry distinct from this one is obtained 
  by exploiting the discrete transformation $R$  which acts on the doublet of fermions:
    \bea
    R
    \left(
    \begin{array}{c}
    \lambda^a   \\
    \psi^a   \\
    \end{array}
    \right) R^{-1}
     =     \left(
           \begin{array}{c}
           \psi^a  \\
         - \lambda^a   \\
           \end{array}
           \right)~.
           \label{discreteR}
    \eea 
 As is explained in the appendix of \cite{FIS1}, we define the second supersymmetry
 by
  \bea
  \label{seconsusydef}
    \delta^{(2, \Im e)}_{\theta}
     \equiv   R \delta^{(1, - \Im e)}_{\theta} R^{-1},
    \eea
  taking the sign flip of $\Im e$ at the action into account.
  The invariance of the action $S( \Im e)$ under this second supersymmetry follows from that of the first one:
  \bea
     \delta^{(2, \Im e)}_{\theta} S(\Im e)=  R \delta^{(1,- \Im e)}_{\theta} R^{-1} R S(- \Im e)  R^{-1}
     =  R \delta^{(1, - \Im e)}_{\theta} S(- \Im e)  R^{-1} =0\;.
  \eea

  We have collected the supersymmetry transformation constructed by eq. (\ref{seconsusydef}) in appendix \ref{sec:susytr}.
  In particular, the supersymmetry transformation acting on the doublet of  fermions (\ref{FIS'susytr}) is
    \bea
    \delta \left(
    \begin{array}{c}
    \lambda^a \\
    \psi^a
    \end{array}
    \right) 
    &=&    F_{\mu \nu}^a \sigma^{\mu \nu} 
           \left( \begin{array}{c}
           \eta_1 \\
           \eta_2
           \end{array} \right)
         - i \sqrt{2} \sigma^\mu 
           \left( \begin{array}{c}
           \bar{\eta}_2 \\
           - \bar{\eta}_1
           \end{array} \right) \CD_\mu \phi^a
         - \frac{i}{2} g^{ab} \fD_b 
           \left( \begin{array}{c}
           \eta_1 \\
           \eta_2
           \end{array} \right)
           \nonumber \\
    & &  + 2 \sqrt{N} g^{ab} \left( \begin{array}{cc}
           0 & e \delta^0_b + m \bar{\CF}_{0b} \\
           - (\bar{e} \delta^0_b + m \bar{\CF}_{0b}) & 0
           \end{array} \right)
           \left( \begin{array}{c}
           \eta_1 \\
           \eta_2
           \end{array} \right)
           \nonumber \\
    & &  + \frac{\sqrt{2}i}{4} g^{ab}
           \left( \begin{array}{cc}
         - (\CF_{bcd} \psi^c \lambda^d + \bar{\CF}_{bcd} \bar{\psi}^c \bar{\lambda}^d),
           & \CF_{bcd} \lambda^c \lambda^d - \bar{\CF}_{bcd} \bar{\psi}^c \bar{\psi}^d \\
         - (\CF_{bcd} \psi^c \psi^d - \bar{\CF}_{bcd} \bar{\lambda}^c \bar{\lambda}^d), 
           & \CF_{bcd} \psi^c \lambda^d + \bar{\CF}_{bcd} \bar{\psi}^c \bar{\lambda}^d
           \end{array} \right)   
           \left( \begin{array}{c}
           \eta_1 \\
           \eta_2
           \end{array}\right),
           \label{susytrfermions}
    \eea
  where $\delta \Psi = \delta^{(1)} \Psi + \delta^{(2)} \Psi$.

\subsection{Some properties of the classical vacua and the mass spectrum}
\label{subsec:FISclassicalvacua}
  In the following, we will analyze the prototypical case of a single trace prepotential of degree $n + 2$:
    \bea
    \CF(\Phi)
     =     \sum_{k = 0}^{n} \frac{\tilde{g}_k}{(k + 2)!} \Tr \Phi^{k + 2}\;.
           \label{prepotential}
    \eea
  With this choice, the superpotential becomes essentially that considered 
  in \cite{CIV, DV, DGLVZ, CDSW}:
    \bea
    W(\Phi)
     =     \Tr \left( 2 e \Phi
         + m \sum_{k = 0}^{n} \frac{\tilde{g}_k}{(k + 1)!} \Phi^{k + 1} \right).
           \label{superpotential}
    \eea
  
  Let us consider the classical vacua of the model, which preserve $\CN=1$ supersymmetry.
  We analyze the scalar potential of this theory (\ref{scalarpotential}):
    \bea
    V
     =     g^{ab}
           \left(
           \frac{1}{8} \fD_a \fD_b 
         + \partial_a W
           \overline{
           \partial_{b} W}
           \right), 
           \label{potential}
    \eea
  where $\fD_a = - i g_{ab} f^b_{cd} \bar{\phi}^c \phi^d$.
  The vacuum  condition is
    \bea
    \partial_a V
    &=&  - \frac{1}{2 i} g^{bd} \CF_{ade} g^{ec} (e \delta_b^0 + m \bar{\CF}_{0b}) 
           (\bar{e} \delta_c^0 + m \bar{\CF}_{0c})
     =     0.
    \eea
  Note that we are considering the vacua $\langle \phi^r \rangle =0$ where $\fD_a$ term vanishes.
  We have decomposed the gauge index $a$ into $a = (i, r)$, where $i$ and $r$ label 
  the Cartan and non-Cartan parts respectively.
  In order to analyze the above conditions, we introduce another basis  
  such that  the elements of the Cartan subalgebra are
  $(t_{\ui})_j^{~k} = \delta_{\ui}^{~k} \delta_j^{~\ui}$ ($\ui = 1, \ldots, N$) \cite{FIS2}.
  In this basis, the vacuum conditions are simply written as
    \bea
    \langle
    \CF_{\ui \ui \ui} (g^{\ui \ui})^2 (2 e + m \bar{\CF}_{\ui \ui})(2 \bar{e} + m \bar{\CF}_{\ui \ui})
    \rangle
     =      0,
            ~~~~
            \ui~{\rm not ~summed},
            \label{cond}
    \eea
  for each $\ui$. 
  As $\langle \CF_{\ui \ui \ui} \rangle = 0$ or $\langle g^{\ui \ui} \rangle = 0$ corresponds to unstable vacua, 
  the above condition reduces to 
  $\langle (2 e + m \bar{\CF}_{\ui \ui})(2 \bar{e} + m \bar{\CF}_{\ui \ui}) \rangle = 0$ for each $\ui$.
  As the parameter $e$ is complex, we have to choose 
    \bea
    \langle \CF_{\ui \ui} \rangle
     =   - \frac{2 e}{m}
             ~~{\rm or}~~
         - \frac{2 \bar{e}}{m},
    \eea 
  for each $\ui$.
  
  Taking this into account, the possible $\CN=1$ supersymmetric vacua are as follows. 
  In the case with $\frac{\Im e}{m} < 0$, we have two possibilities. 
  The first one is where $\langle \CF_{\ui \ui} \rangle = - \frac{2 e}{m}$ for all $\ui$.
  (In the original bases, $m \langle \CF_{00} \rangle = - e$.)
  These vacua  are just obtained from the F-term equation: these vacua preserve 
  the $\CN=1$ supersymmetry which is manifest in the Lagrangian (\ref{FIS'}).
  The vacuum expectation value of $\CF$ determines the vacuum value of the scalar field.
  As a result, the gauge symmetry is  broken  in general into
  $ U(N) \rightarrow \prod_{i=1}^n U(N_i)$.
  As can be seen from (\ref{susytrfermions}), the NGF associated with the partial supersymmetry breaking
  is $\lambda^0$ residing in the overall $U(1)$ part.
  
  While we do not treat in this paper, there is another possibility 
  that $\langle \CF_{\ui \ui} \rangle = - \frac{2 \bar{e}}{m}$ for all $\ui$.
  (In the original bases, $\langle \CF_{00} \rangle = - \frac{\bar{e}}{m}$).
  In contrast to the first vacua, these vacua preserve another $\CN=1$ supersymmetry, as analyzed in \cite{FIS2}.
  Actually, we can see from (\ref{susytrfermions}) that the NGF is $\psi^0$
  and the first $\CN=1$ supersymmetry is broken.
  Note that $\langle g_{\ui \ui} \rangle$ are all negative in these vacua.
  Although this leads to the negative kinetic energy, 
  it has been argued in \cite{ABSV, Maruyoshi} that there exists a well-defined description.
  In the case with $\frac{\Im e}{m} > 0$, the situation is the opposite to that of the above \cite{FIS2, Maruyoshi}.

\begin{figure}
     \begin{center}
     \includegraphics[scale=0.8]{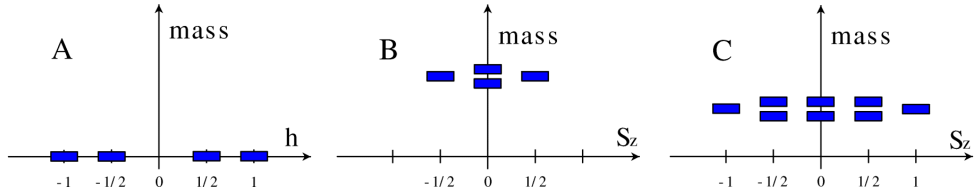}
     \caption{{\small The mass spectrum.}}
     \label{fig:NGFmass}
     \end{center}
     \end{figure}

  Let us now turn to the discussion of the mass spectrum of the model. 
  The mass spectrum was derived in \cite{FIS2} and we here briefly recall some of the qualitative features. 
  There are three types of unbroken $\CN =1$ multiplets which we refer to as type A, type B and type C
  and are depicted in Fig.~\ref{fig:NGFmass}.
  Type A supermultiplet is massless and consists of two polarization states of helicity $1/2$ and $1$ 
  and their CPT conjugate.
  The NG supermultiplet lies in the overall $U(1)$.
  Type B supermultiplet consists of massive states of spin ($z$ component) $1/2$, $-1/2$ and two of spin zero.
  This supermultiplet receives  the mass of 
  $| \sqrt{2 N} m \langle g^{\alpha \alpha} \CF_{0 \alpha \alpha} \rangle |$
  through the third prepotential derivatives, which is our characteristic mass generation mechanism. 
  We have defined the indices as $\alpha = \{ \ui, r | [t_r, \langle \phi \rangle] =0 \}$.
  ($r$ label the non-Cartan generators corresponding to the unbroken gauge symmetry.
  The corresponding generators $t_r$ can be written as $E_{\ui \uj}^{\pm}$ 
  with $\langle \phi^{\ui} \rangle = \langle \phi^{\uj} \rangle$.
  Here $E_{\ui \uj}^{+} = \frac{1}{2} (E_{\ui \uj} + E_{\uj \ui})$ 
  and $E^-_{\ui \uj} = - \frac{i}{2} (E_{\ui \uj} - E_{\uj \ui})$.
  $E_{\ui \uj}$ has the nonvanishing entry $1$ at the ($\ui, \uj$) element only.)
  It is a salient feature of the model that mass is naturally supplied to the scalar this way.
  The scalar of this kind has received attention recently 
  and is called s-gluon in some of the phenomenological researches \cite{ChoiDrees}.
  Type C supermultiplet consists  of a set of polarization states of spin $(0, 1/2, 1/2, 1)$ and its CPT conjugate.
  This supermultiplet receives its mass through the Higgs mechanism.
  
\subsection{Nondecoupling of $U(1)$ from $SU(N)$ and NGF-related vertices}
\label{subsec:vertex}
  In this section, we will see that the overall $U(1)$ gauge part (in which the NGF resides) does not decouple 
  from the other part in the lagrangian.
  To see this we concentrate on the Yukawa interaction terms.
  
  As in appendix \ref{sec:component}, 
  the Yukawa couplings are contained in $\CL_{\rm{mass}}$ (\ref{Lmass}) and $\CL_{\rm{kin}}$ (\ref{Lkin}).
  Let us write them down for convenience:
    \bea
    \CL_{\rm{mass}}
    &=&  - \frac{\sqrt{2N}}{2} m \CF_{0ab} \psi^a \psi^b 
         - \frac{i \sqrt{2N}}{4} g^{ab} (e \delta^0_a + m \CF_{0a}) \CF_{bcd} \psi^c \psi^d
           \label{Lmass'} \\
    & &  - \frac{i \sqrt{2N}}{4} g^{ab} (\bar{e} \delta^0_a + m \bar{\CF}_{0a}) \CF_{bcd} \lambda^c \lambda^d
         + \frac{1}{\sqrt{2}} g_{ab} f^b_{cd} \bar{\phi}^d \lambda^c \psi^a
         - \frac{1}{4 \sqrt{2}} f^a_{bc} \bar{\phi}^b \phi^c \CF_{ade} \psi^d \lambda^e
         + c.c.,
           \nonumber \\
    \CL_{\rm{kin}}
    &=&  - \frac{1}{2} \CF_{ab} \lambda^a \sigma^\mu \mathcal{D}_\mu \bar{\lambda}^b
         - \frac{i}{2} g_{ab} \psi^a \sigma^\mu \mathcal{D}_\mu \bar{\psi}^b
         + c.c. + \ldots,
           \label{Lkin'}
    \eea
  where we have substituted the explicit form of the superpotential (\ref{superpotential}) into (\ref{Lmass}).
  
\begin{figure}
     \begin{center}
     \includegraphics[scale=0.65]{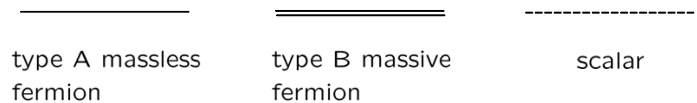}
     \caption{{\small Feynman propagators for the massless and massive fermions and the scalar.
                      That of type A massless fermions, i.e., $\lambda^a$ is drawn by a single solid line 
                      while that of type B massive fermions, i.e., $\psi^a$ is drawn by a double solid lines. 
                      A scalar propagator is drawn by a broken line.}}
     \label{fig:NGFpropa}
     \end{center}
     \end{figure}

\begin{figure}
     \begin{center}
     \includegraphics[scale=0.8]{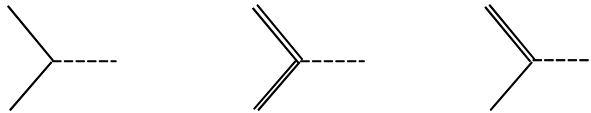}
     \caption{{\small A Feynman rule for three-point vertices
                      corresponding to (\ref{psipsiphi}), (\ref{lambdalambdaphi}) and (\ref{psilambdaphi}).}}
     \label{fig:NGF3point}
     \end{center}
     \end{figure}

  Let us consider the model on the vacua with partially and spontaneously broken supersymmetry. 
  We expand the scalar fields around the vacuum expectation values as
    \begin{equation}
    \phi^a
     =     \langle \phi^a \rangle + \tilde{\phi}^{a}.
           \label{vev1}
    \end{equation}
  We expand the prepotential and other quantities, e.g., 
  $ \mathcal{F}_{abc} = \langle \mathcal{F}_{abc} \rangle + \langle \mathcal{F}_{abcd} \rangle \tilde{\phi}^d + \dots$
  in the fluctuation field $\tilde{\phi}$,
  
  Let us list the Yukawa couplings obtained from $\CL_{{\rm mass}}$ (\ref{Lmass'}).
  From the first and second terms, the $\psi \psi \phi$ vertices are 
    \bea
    \frac{\sqrt{2N}}{4} m \langle - 2i \CF_{0abc} + g^{de} \CF_{abe} \CF_{0cd} \rangle
    \tilde{\phi}^c \psi^a \psi^b.
    \label{psipsiphi}
    \eea
  Note that we have included $i$ factor in front of the interaction lagrangian.
  The $\lambda \lambda \bar{\phi}$ vertices are calculated from the third term of (\ref{Lmass'}):
    \bea
    \frac{\sqrt{2N}}{4} m \langle g^{de} \CF_{abe} \bar{\CF}_{0cd} \rangle \bar{\tilde{\phi}}^c \lambda^a \lambda^b.
    \label{lambdalambdaphi}
    \eea
  From the last two terms, we obtain the $\lambda \psi \phi$ ($\lambda \psi \bar{\phi}$) vertices
    \bea
    & &    \frac{1}{4 \sqrt{2}} \langle 2 f^b_{cd} \CF_{abe} \bar{\phi}^d 
         - f^b_{de} \CF_{abc} \bar{\phi}^d \rangle 
           \tilde{\phi}^e \lambda^c \psi^a
           \nonumber \\
    & &  + \frac{1}{4 \sqrt{2}} \langle 4 i f^b_{ce} g_{ab} - 2 f^b_{cd} \bar{\CF}_{abe} \bar{\phi}^d 
         - f^b_{ed} \CF_{abc} \phi^d \rangle 
           \bar{\tilde{\phi}}^e \lambda^c \psi^a.
           \label{psilambdaphi}
    \eea
  There are also  derivative coupling terms obtained from $\CL_{{\rm kin}}$ (\ref{Lkin'}).
  The Feynman rules are illustrated in Fig.~\ref{fig:NGFpropa} and \ref{fig:NGF3point}.
  The propagator of the massless $\lambda$ fermions and that of massive $\psi$ fermions are denoted 
  by a single line and a double solid line respectively. 
  The scalar propagator is drawn by a broken line.
    
  Next, we focus on the Yukawa coupling involving one NGF.
  Such couplings can be obtained from $\lambda \lambda \phi$ and $\psi \lambda \phi$ vertices in $\CL_{{\rm mass}}$.
  From (\ref{lambdalambdaphi}), it can be easily seen 
  that the coupling of the NGF and the $SU(N)$ fermions are indeed nonvanishing:
    \bea
    \frac{\sqrt{2N}}{4} m \langle g^{de} \CF_{0be} \bar{\CF}_{0cd} \rangle \bar{\tilde{\phi}}^c \lambda^0 \lambda^b
     =     \frac{\sqrt{2N}}{4} m \sum_{\alpha} \langle g^{\alpha \alpha} |\CF_{0 \alpha \alpha}|^2 \rangle 
           \tilde{\phi}^\alpha \lambda^0 \lambda^\alpha.
           \label{lambdalambdaphiNGF}
    \eea
  Note that we have used $\langle \CF_{0 \mu \nu} \rangle = 0$ \cite{FIS2},
  where $\mu$ label the broken non-Cartan generators.
  Eq. (\ref{lambdalambdaphiNGF}) means that the Yukawa coupling of the NGF with 
  the fermions belonging to the unbroken generators {\it does} exist.
  The existence of the {\it nonderivative} Yukawa coupling that involves the NGF and that is supported 
  by the third prepotential derivatives is remarkable.
  
  The coupling coming from (\ref{psilambdaphi}) vanishes 
  because $f^0_{ab} = 0$ and $f^{r}_{ab} \langle \phi^a \rangle = 0$ which follows from $[t_r , \langle \phi \rangle] = 0$.
  There are also derivative couplings due to $\CL_{{\rm kin}}$.

\section{Low energy suppression of NGF emission by direct computation}
\label{sec:LEsuppression}
  In this section, we check the validity of the low energy theorem stated in section \ref{sec:LEtheorem} 
  by direct computation of tree Feynman diagrams obtained from the lagrangian. 
  According to our discussion of the $U(1)$ nondecoupling 
  and the presence of nonderivative couplings of the NGF and the $SU(N)$ sector, 
  the low energy suppression of processes having the emission of the NGF 
  together with massless as well as massive particles in the final state is by no means obvious. 
  We will exhibit a cancellation mechanism shortly.
  
\begin{figure}
     \begin{center}
     \includegraphics[scale=0.75]{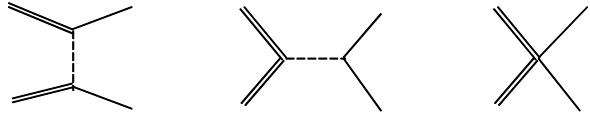}
     \caption{{\small $\psi^0 \psi^\alpha \rightarrow \lambda^0 \lambda^{\alpha}$ tree diagrams.}}
     \label{fig:NGF4point}
     \end{center}
     \end{figure}

  For simplicity and illustrative purposes, 
  we take as an initial state two massive fermions of type B with momentum $p^a$ and $p^b$.  
  In the final state, we consider the case in which the NGF with momentum $p^{0}$ and massless fermion (gaugino) 
  of type A with momentum $p^{\alpha}$ are present. 
  We limit ourselves to this case, namely, $\psi^a \psi^b \rightarrow \lambda^0 \lambda^{\alpha}$ scattering in this paper,
  as in Fig.\ref{fig:NGF4point}.
  
  The nonvanishing possibilities are $(a,b)= (\alpha, 0)$.
  As for the diagram of $t$-channel scalar exchange,
  the relevant interaction vertex can in principle be obtained from (\ref{psilambdaphi}).
  But, as explained in subsection \ref{subsec:vertex},
  it vanishes for the NGF as the piece of the structure constant is zero in which overall $U(1)$ is involved.
  
  In order to discuss the diagram of $s$-channel annihilation,
  it is more transparent to extract the appropriate effective $(\psi \psi)(\lambda \lambda)$
  vertex with one scalar contraction from $\frac{1}{2} \left( i \int \CL_{{\rm mass}} \right)^2$.
  Rescaling the fluctuation field as $\tilde{\phi}^a = \sqrt{g^{aa}} \phi^a_{can}$, we obtain
    \bea
    \label{llpsps}
    & &    \frac{1}{16} (\sqrt{2N} m)^2 \langle g^{ab} \bar{\CF}_{0be} \sqrt{g^{ee}} \CF_{acd} \rangle
           \langle g^{hi} \CF_{0hf} \sqrt{g^{ff}} \CF_{ijk} 
         - 2i \CF_{0hif} \sqrt{g^{ff}} \rangle
           \\
    & &    ~~~~~~
           \times \int d^4 x_1 d^4 x_2 \langle \phi^{* e}_{can }(x_1) \phi^f_{can }(x_2) \rangle 
           (\lambda^c \lambda^d)(x_1)(\psi^j \psi^k)(x_2)\;.
           \nonumber 
    \eea
  Note that
    \bea
    {\rm F.T.} \left( \langle \phi^{* e}_{can }(x_1) \phi^f_{can }(x_2) \rangle \right) (p_0, p_\alpha)
     =     \frac{-i \delta_{ef}}{m_{\alpha}^2 - (p_0 + p_{\alpha})^2}
    \eea  
  as well as
    \bea
    m_{\alpha}^2
     =     2 N m^2 \langle g^{\alpha \alpha} \rangle^2 \langle \CF_{0 \alpha \alpha} \rangle 
           \langle \bar{\CF}_{0 \alpha \alpha}\rangle \;.
    \eea
  In the limit of $p_0 \rightarrow 0$, the propagator is $\frac{-i}{m_{\alpha}^2}$, 
  and the contribution to the $\psi^a \psi^b \rightarrow \lambda^0 \lambda^{\alpha}$ scattering amplitude 
  from eq.~(\ref{llpsps}) is
    \bea
    \label{s^annlim}
     - \frac{1}{8} \langle \CF_{00 \alpha \alpha} \rangle 
     - \frac{i}{16} \langle g^{\alpha \alpha} \rangle \langle \CF_{0 \alpha \alpha} \rangle^2 \;.
    \eea
  The presence of this alone is against the low energy theorem and is in fact saved by
  the presence of the appropriate four-Fermi interactions in (A.18). 
  Their contributions are
    \bea
    \frac{1}{8} \langle \CF_{0 0 \alpha \alpha} \rangle + \frac{i}{16} \langle g^{\alpha \alpha} \rangle
    ( \langle \CF_{0 \alpha \alpha} \rangle)^2 \;,
    \eea
   exactly cancelling eq.~(\ref{s^annlim}).
  
\section*{Acknowledgements}
  We would like to thank Taichiro~Kugo for useful comments.
  The research of H.~I.~ is supported in part by the Grant-in-Aid for Scientific Research (2054278) 
  from the Ministry of Education, Science and Culture, Japan.
  The research of K.~M.~ is supported in part by JSPS Research Fellowships for Young Scientists. 
  H.~I.~ acknowledges the hospitality of YITP extended to him
  during the period of the workshop ``Branes, Strings and Black Holes."

\appendix

\section*{Appendix}
\section{The component Lagrangian}
\label{sec:component}
  In this appendix, we consider the lagrangian of $\CN=2$ supersymmetric $U(N)$ gauge theory 
  with the electric and magnetic FI terms in terms of the component fields.
  We use convention of \cite{Lykken}.
  In particular, we use the metric $\eta^{\mu \nu} = \diag(+ 1, - 1, -1 , -1)$
  and the Levi-Civita symbol $\epsilon_{0123} = +1$ and $\epsilon^{0123} = -1$.
  
  The lagrangian (\ref{FIS'}) can be divided into the following parts:
    \bea
    \CL
     =     \CL_{{\rm Kahler}} + \CL_{{\rm gauge}} + \left[ \int d^2 \theta W + c.c. \right],
    \eea
  where
    \bea
    \CL_{{\rm Kahler}}
    &=&    \Im 
           \left[ 
           \int d^4 \theta 
           {\rm Tr} 
           \bar{\Phi} e^{ad V} 
           \frac{\partial \CF(\Phi)}{\partial \Phi} \right],
           \nonumber \\
    \CL_{{\rm gauge}}
    &=&    \Im 
           \left[ \int d^2 \theta
           \frac{1}{2} 
           \CF_{ab}
           \CW^{\alpha a} \CW^b_{\alpha}
           \right],
    \eea
  We have chosen the common function $\CF$ in K\"ahler, gauge kinetic and superpotential terms
  such that the lagrangian is invariant under the discrete R transformation (\ref{discreteR}).
  
  In components, as can be seen in appendix of \cite{IM2}, 
  $\CL_{{\rm Kahler}}$ is the same as $\CL_K + \CL_\Gamma$ in the original lagrangian of \cite{FIS1},
  which is
    \bea
    \CL_{{\rm Kahler}}
    &=&    g_{ab}\CD_\mu \phi^a \CD^\mu \bar{\phi}^b
         - \frac{i}{2} g_{ab} \psi^a \sigma^\mu \CD_\mu' \bar{\psi}^b
         + \frac{i}{2} g_{ab} \CD_\mu' \psi^a \sigma^\mu \bar{\psi}^b + g_{ab} F^a \bar{F}^b
           \nonumber \\
    & &  - \frac{1}{2} g_{ab,\bar{c}} F^a \bar{\psi}^b \bar{\psi}^c
         - \frac{1}{2} g_{bc,a} \bar{F}^c \psi^a \psi^b
         + \frac{1}{\sqrt{2}} g_{ab} (\lambda^c \psi^a k_c^*{}^{b} + \bar\lambda^c\bar\psi^bk_c{}^{a})
         + \frac{1}{2} D^a \fD_a~,
           \label{L;K}
    \eea
  where $g_{ab}$ is the K\"ahler metric and its derivatives are defined as 
  $g_{ab,c} \equiv \partial g_{ab}/ \partial \phi^c$ 
  and $g_{ab,\bar{c}} \equiv \partial g_{ab}/ \partial \bar{\phi}^c$.
  The covariant derivatives are defined as
    \bea
    \CD_\mu \phi^a
    &=&    \partial_\mu \phi^a - \frac{1}{2} A_\mu^b k_b{}^a,
           \\
    \CD_\mu' \psi^a
    &=&    \CD_\mu \psi^a + \Gamma^a_{bc} (\CD_\mu \phi^b) \psi^c,
           \\
    \CD_\mu \psi^a
    &=&    \partial_\mu \psi^a
         - \frac{1}{2} A_\mu^b \partial_c k_b{}^a \psi^c~,
           \label{D psi}
    \eea
  where $\Gamma^a_{bc} = g^{ad} g_{bd,c}$.
  Also, the Killing vector and the Killing potential are given by
    \bea
    k_a 
    &=&    k_a^b \partial_b,
           ~~~
    k_a^b
     =   - i g^{bc} \bar{\partial}_c \fD_a,
           \nonumber \\
    \fD_a
    &=&  - \frac{1}{2} (\CF_b f^b_{ac} \bar{\phi}^c + \bar{\CF}_b f^b_{ac} \phi^c),
           \label{killingpot}
    \eea
  which satisfies \cite{FIS1}
    \bea
    k_b^c \partial_c \Phi^a
     =     f^a_{bc} \Phi^c,
           ~~~
    k_b^c \partial_c \CF_a
     =   - f^a_{bc} \CF_a.
           \label{kandf}
    \eea
  By using the second equation of (\ref{kandf}): $\CF_b f_{ac}^b = k_c^d \partial_d \CF_a = \CF_{ad} f^d_{ce} \phi^e$, 
  the Killing potential (\ref{killingpot}) can be written as
    \bea
    \fD_a
     =   - \frac{1}{2} f^b_{cd} \bar{\phi}^c \phi^d (\CF_{ab} - \bar{\CF}_{ab})
     =   - i g_{ab} f^b_{cd} \bar{\phi}^c \phi^d.
           \label{killingpot2}
    \eea
    
  The gauge part $\CL_{{\rm gauge}}$ is, in components,
    \bea
    \CL_{{\rm gauge}}
    &=&  - \frac{1}{2} \CF_{ab} \lambda^a \sigma^\mu \CD_\mu \bar{\lambda}^b
         - \frac{1}{2} \bar{\CF}_{ab} \CD_\mu \lambda^a \sigma^\mu \bar{\lambda}^b
         - \frac{1}{4} g_{ab} F_{\mu \nu}^a F^{b \mu \nu}
         - \frac{1}{8} (\Re \CF)_{ab} \epsilon^{\mu \nu \rho \sigma} F_{\mu \nu}^a F_{\rho \sigma}^b
           \nonumber \\
    & &  - \frac{\sqrt{2} i}{8} (\CF_{abc} \psi^c \sigma^\nu \bar{\sigma}^\mu \lambda^a
         - \bar{\CF}_{abc} \bar{\lambda}^a \bar{\sigma}^\mu \sigma^\nu \bar{\psi}^c ) F_{\mu \nu}^b
           \nonumber \\
    & &  + \frac{1}{2} g_{ab} D^a D^b
         + \frac{\sqrt{2}}{4} ( \CF_{abc} \psi^c\lambda^a
         + \bar{\CF}_{abc} \bar{\psi}^c \bar{\lambda}^a ) D^b
         + \frac{i}{4} \CF_{abc} F^c \lambda^a \lambda^b
         - \frac{i}{4} \bar{\CF}_{abc} \bar{F}^c \bar{\lambda}^a \bar{\lambda}^b
           \nonumber \\
    & &  - \frac{i}{8} \CF_{abcd} \psi^c \psi^d \lambda^a \lambda^b
         + \frac{i}{8} \bar{\CF}_{abcd} \bar{\psi}^c \bar{\psi}^d \bar{\lambda}^a \bar{\lambda}^b,
           \label{L:gauge}
    \eea
  where the field strength is 
  $F_{\mu \nu}^a = \partial_\mu A_\nu^a - \partial_\nu A_\mu^a - \frac{1}{2} f^a_{bc} A_\mu^b A_\nu^c$.
  Finally, the superpotential can be written as
    \bea
    \int d^2 \theta W(\Phi) +c.c.
    &=&    F^a \partial_a W
         - \frac{1}{2} \partial_a \partial_b W \psi^a \psi^b
         + c.c..
           \label{L:superpotential}
    \eea
  
  Let us exhibit the on-shell Lagrangian.
  Eq. of motion with respect to the auxiliary fields $D^a$ and $F^a$ are
    \bea
    D^{a}
    &=&  - \frac{1}{2} g^{ab} \fD_b
         - \frac{1}{2 \sqrt{2}} g^{ab} 
           \left( \CF_{bcd}\psi^d \lambda^c + \bar{\CF}_{bcd} \bar{\psi}^d \bar{\lambda}^c \right), 
           \nonumber \\
    F^a
    &=&  - g^{ab} \overline{\partial_{b} W}
         - \frac{i}{4} g^{ab} \left( \CF_{bcd} \psi^c \psi^d - \bar{\CF}_{bcd} \bar{\lambda}^c \bar{\lambda}^d \right),
           \label{DFF}
    \eea
  where $g^{ab}$ are defined by $g_{ab} g^{bc} = \delta^c_a$.
  After eliminating the auxiliary fields, the lagrangian $\CL$ reduces to the following on-shell Lagrangian:
    \bea
    \label{action:on-shell}
    \CL_{{\rm on-shell}}
    &=&    \CL_{\rm{kin}} + \CL_{\rm{pot}} + \CL_{\rm{Pauli}} + \CL_{\rm{mass}} + \CL_{\rm{fermi^4}},
    \eea
  where
    \bea
    \CL_{\rm{kin}}
    &=&    g_{ab} \mathcal{D}_\mu \phi^a \mathcal{D}^\mu \bar{\phi}^b
         - \frac{1}{4} g_{ab} F_{\mu \nu}^a F^{b \mu \nu}
         - \frac{1}{8} (\Re \CF)_{ab} \epsilon^{\mu \nu \rho \sigma} F_{\mu \nu}^a F_{\rho \sigma}^b
           \label{Lkin} \\
    & &  - \frac{1}{2} \CF_{ab} \lambda^a \sigma^\mu \mathcal{D}_\mu \bar{\lambda}^b
         - \frac{1}{2} \bar{\CF}_{ab} \mathcal{D}_\mu \lambda^a \sigma^\mu \bar{\lambda}^b
         - \frac{i}{2} g_{ab} \psi^a \sigma^\mu \mathcal{D}_\mu \bar{\psi}^b
         + \frac{i}{2} g_{ab} \mathcal{D}_\mu \psi^a \sigma^\mu \bar{\psi}^b,
           \nonumber \\
    \CL_{\rm{pot}}
    &=&  - \frac{1}{8} g^{ab} \fD_{a} \fD_{b} 
         - g^{ab} \partial_a W \overline{\partial_b W} 
     =   - V,
           \label{scalarpotential}
           \\
    \CL_{\rm{Pauli}}
    &=&  - \frac{i}{4 \sqrt{2}} \CF_{abc} \psi^c \sigma^\nu \bar{\sigma}^\mu \lambda^a F_{\mu \nu}^b
         + \frac{i}{4 \sqrt{2}} \bar{\CF}_{abc} \bar{\lambda}^a \bar{\sigma}^\mu \sigma^\nu \bar{\psi}^c F_{\mu \nu}^b,
           \\
    \CL_{\rm{mass}}
    &=&  - \frac{1}{2} \partial_a \partial_b W \psi^a \psi^b 
         - \frac{i}{4} g^{ab} \partial_a W 
           \left( \CF_{bcd} \psi^c \psi^d - \bar{\CF}_{bcd} \bar{\lambda}^c \bar{\lambda}^d \right)
           \nonumber \\
    & &  - \frac{1}{2} \overline{\partial_a \partial_b W} \bar{\psi}^a \bar{\psi}^b
         - \frac{i}{4} g^{ab} \overline{\partial_a W}
           \left( \CF_{bcd} \lambda^c \lambda^d - \bar{\CF}_{bcd} \bar{\psi}^c \bar{\psi}^d \right)
           \nonumber \\
    & &  + \frac{1}{\sqrt{2}} g_{ab} \left( \bar{\lambda^c} \bar{\psi}^b k_c{}^a + \lambda^c \psi^a k_c^*{}^{b} \right)
         - \frac{1}{4 \sqrt{2}} g^{ab} \fD_{a}
           \left( \CF_{bcd} \psi^d \lambda^c + \bar{\CF}_{bcd} \bar{\psi}^d \bar{\lambda}^c \right),
           \label{Lmass}
           \\
    \CL_{\rm{fermi^4}}
    &=&  - \frac{i}{8} \CF_{abcd} \psi^a \psi^b \lambda^c \lambda^d
         + \frac{i}{8} \bar{\CF}_{abcd} \bar{\psi}^a \bar{\psi}^b \bar{\lambda}^c \bar{\lambda}^d
           \nonumber \\
    & &  - \frac{1}{16} g^{ab}
           \left( \CF_{acd} \psi^d \lambda^c + \bar{\CF}_{acd} \bar{\psi}^d \bar{\lambda}^c \right)
           \left( \CF_{bef} \psi^f \lambda^e + \bar{\CF}_{bef} \bar{\psi}^f \bar{\lambda}^e \right)
           \nonumber \\
    & &  + \frac{1}{16} g^{ab}
           \left( \CF_{acd} \lambda^c \lambda^d - \bar{\CF}_{acd} \bar{\psi}^c \bar{\psi}^d \right)
           \left( \CF_{bef} \psi^e \psi^f - \bar{\CF}_{bef} \bar{\lambda}^e \bar{\lambda}^f \right).
           \label{L4fermi}
    \eea
  In (\ref{scalarpotential}), we have used that $\fD_a g^{ab} \delta_b^0 = - i f^0_{ab} \bar{\phi}^a \phi^b = 0$,
  following from (\ref{killingpot2}).
  
\section{Supersymmetry transformation law}
\label{sec:susytr}
  We consider supersymmetry transformation laws in this subsection.
  The first and second $\CN=1$ supersymmetry transformation laws of the scalar and the fermions are 
  (see \cite{Lykken}):
    \bea
    \delta_{\eta_1} \phi^a
    &=&    \sqrt{2} \eta_1\psi^a,
           \nonumber \\
    \delta_{\eta_1} \psi^a
    &=&    i \sqrt{2}\sigma^\mu \bar{\eta}_1 \CD_\mu \phi^a + \sqrt{2} \eta_1 F^a, 
           \nonumber \\
    \delta_{\eta_1} \lambda^a
    &=&    \sigma^{\mu \nu} \eta_1 F_{\mu \nu}^a + i \eta_1 D^a,
           \nonumber\\
    \delta_{\eta_1} A_\mu^a
    &=&  - i \eta_1 \sigma_\mu \bar{\lambda}^a + i \lambda^a \sigma_\mu \bar{\eta}_1,
           \label{1stsusy}
    \eea
  and
    \bea
    \delta_{\eta_2} \phi^a
    &=&  - \sqrt{2} \eta_2 \lambda^a,
           \nonumber \\
    \delta_{\eta_2} \lambda^a
    &=&  - i \sqrt{2} \sigma^\mu \bar{\eta}_2 \CD_\mu \phi^a - \sqrt{2} \eta_2 \widetilde{F}^a, 
           \nonumber \\
    \delta_{\eta_2} \psi^a
    &=&    \sigma^{\mu \nu} \eta_2 F_{\mu \nu}^a + i \eta_2 \widetilde{D}^a,
           \nonumber \\
    \delta_{\eta_2} A_\mu^a
    &=&  - i \eta_2 \sigma_\mu \bar{\psi}^a + i \psi^a \sigma_\mu \bar{\eta}_2,
           \label{2ndsusy}
    \eea
  where $\eta_1$ and $\eta_2$ are the transformation parameters 
  of the first $\CN=1$ supersymmetry and the second one respectively.
  The second supersymmetry transformation is derived by acting the discrete R transformation 
  on the first one, as explained in section \ref{sec:lagrangian}.
  Also, the auxiliary fields are defined in (\ref{DFF}) and
    \bea
    \widetilde{D}^{a}
    &=&  - \frac{1}{2} g^{ab} \fD_b
         + \frac{1}{2 \sqrt{2}} g^{ab} 
           \left( \CF_{bcd}\psi^d \lambda^c + \bar{\CF}_{bcd} \bar{\psi}^d \bar{\lambda}^c \right), 
           \nonumber \\
    \widetilde{F}^a
    &=&  - \sqrt{2N} g^{ab} (e \delta_b^0 + m \bar{\CF}_{0b})
         - \frac{i}{4} g^{ab} \left( \CF_{bcd} \lambda^c \lambda^d - \bar{\CF}_{bcd} \bar{\psi}^c \bar{\psi}^d \right). 
    \eea
  We note that the sign of $\Im e$ has been flipped in $\widetilde{F}$ as compared with $F$ in (\ref{DFF}).
  
  These supersymmetry transformation laws (\ref{1stsusy}) and (\ref{2ndsusy}) can be combined into the following
   forms:
    \bea
    \delta \phi^a
    &=&    \sqrt{2} 
           \left(
           \begin{array}{cc}
           \psi^a & - \lambda^a
           \end{array}
           \right)
           \left(
           \begin{array}{c}
           \eta_1 \\
           \eta_2
           \end{array}
           \right),
           \nonumber \\
    \delta \left(
    \begin{array}{c}
    \lambda^a \\
    \psi^a
    \end{array}
    \right) 
    &=&    F_{\mu \nu}^a \sigma^{\mu \nu} 
           \left(
           \begin{array}{c}
           \eta_1 \\
           \eta_2
           \end{array}
           \right)
         - i \sqrt{2} \sigma^\mu 
           \left(
           \begin{array}{c}
           \bar{\eta}_2 \\
           - \bar{\eta}_1
           \end{array}
           \right) \CD_\mu \phi^a
         + \left( \begin{array}{cc}
           i D^a & - \sqrt{2} \widetilde{F}^a \\
           \sqrt{2} F^a & i \widetilde{D}^a
           \end{array} \right)   
           \left( \begin{array}{c}
           \eta_1 \\
           \eta_2
           \end{array} \right),
           \nonumber \\
    \delta A_\mu^a
    &=&    i \left( \begin{array}{cc}
           \bar{\lambda}^a & \bar{\psi}^a
           \end{array} \right)
           \bar{\sigma}_\mu
           \left( \begin{array}{c}
           \eta_1 \\
           \eta_2
           \end{array} \right)
         + i \left(
           \begin{array}{cc}
           \psi^a & - \lambda^a
           \end{array}
           \right)
           \sigma_\mu
           \left( \begin{array}{c}
           \bar{\eta}_2 \\
           - \bar{\eta}_1
           \end{array} \right)
           \label{FIS'susytr}
    \eea
  where $\delta \Psi \equiv \delta_{\eta_1} \Psi + \delta_{\eta_2} \Psi$.
  The last term of the transformation law of the fermion doublet can be separated into the terms
  involving the fermion bilinears and the other terms.
  This last part is 
    \bea 
    - \frac{i}{2} g^{ab} \fD_b
           \left( \begin{array}{c}
           \eta_1 \\
           \eta_2
           \end{array} \right)
    + 2 \sqrt{N} g^{ab} \left(
           \begin{array}{cc}
           - i \xi & (e \delta_b^0 + m \bar{\CF}_{0b}) \\
           - (\bar{e} \delta_b^0 + m \bar{\CF}_{0b}) & i \xi
           \end{array} \right)   
           \left( \begin{array}{c}
           \eta_1 \\
           \eta_2
           \end{array} \right).
           \label{FIS'susytrfermions}
    \eea
  
\section{$\CN=1$ supercurrent and the low energy theorem}
\label{sec:N=1}
  In this appendix, we briefly review the low energy theorem for the scattering amplitudes 
  in the case where $\CN=1$ supersymmetry is spontaneously and completely broken \cite{dWF}.
  We consider the special case where supersymmetry is broken by the Fayet-Iliopoulos D-term \cite{FI}.
  The supercurrent in the four component majorana notation reads
    \bea
    \CS^\mu
     =   - i f_\lambda \gamma^\mu \lambda 
         + \frac{1}{2} F_{\nu \rho} \gamma^\nu \gamma^\rho \gamma^\mu \gamma_5 \lambda
         + \ldots
           \label{supercurrentN=1}
    \eea
  where $\lambda$ is the NGF field and $f_\lambda$ is the decay constant.
  $F_{\mu \nu}$ is a field strength of an abelian gauge field.
  
  Let us look at the following scattering processes: 
  (1) processes of the type $A \rightarrow B + \lambda$; 
  (2) radiative processes of the type $A \rightarrow B + \gamma + \lambda$.
  We denote the corresponding amplitudes by $\bar{u}(q) M_{1}(q)$ and 
  $\epsilon^*_\mu(k) \bar{u}(q) M_{2}^\mu (k, q)$ respectively.
  $A$ and $B$ are respectively the initial and the final multiparticle states consisting of massive particles alone.
  
\begin{figure}
    \begin{center}
    \includegraphics[scale=0.6]{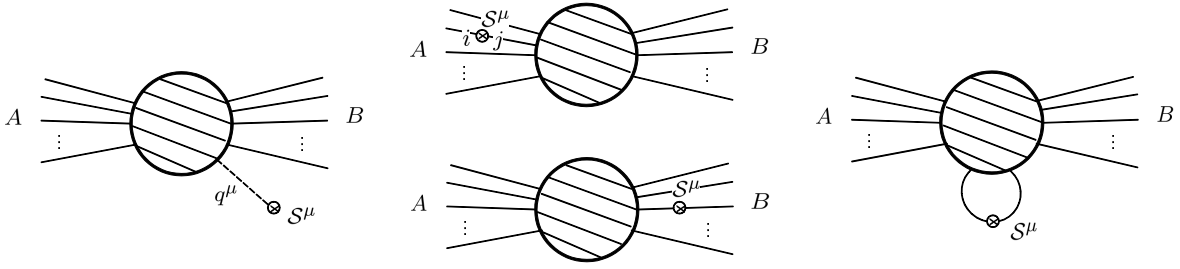}
    \caption{{\small Insertions of $\CS^\mu$ which are denoted by a crossed circle.
                     Fig.~\ref{fig:NGF1}-1) 
                     $\CS^\mu$ is inserted in the end point of the NGF propagator (one-point coupling).
                     Fig.~\ref{fig:NGF1}-2) 
                     $\CS^\mu$ is inserted in an initial or  a final external line (two-point coupling).
                     Fig.~\ref{fig:NGF1}-3) $\CS^\mu$ is connected to internal lines.}}
     \label{fig:NGF1}
     \end{center}
     \end{figure}

  (1) First, consider the matrix element of the form
    \bea
    \langle B | \CS^\mu | A \rangle .
    \eea
  In the diagrammatic representation, the possible insertions of the supercurrent $\CS^\mu$ 
  are divided into three patterns as depicted in Fig.~\ref{fig:NGF1}.
  In the limit $q^{\mu} \rightarrow 0$, the contributions to the amplitude 
  from Fig.~\ref{fig:NGF1}-1 and from Fig.~\ref{fig:NGF1}-2
  can be singular and that from Fig.~\ref{fig:NGF1}-3 is regular.
  In fact, the contribution from Fig.~\ref{fig:NGF1}-1, in terms of the amplitude $M_{1}$ above, reads
    \bea
    \langle 0 | \CS^\mu | \lambda \rangle \times \frac{i q^\mu \gamma_\mu}{q^2} \times M_{1}(q)\;,
    \eea
  which is singular as $q^{\mu} \rightarrow 0$.
  Fig.~\ref{fig:NGF1}-2 can be singular in the limit as well 
  for the case in which masses of the two particles coupling to the current are degenerate.
  This is seen as follows: suppose that we insert $\CS^\mu$ in an initial external line 
  whose momentum and mass are $p_i$ and $m_i$ respectively.
  The propagator that connects to this line via the current insertion is, for the case of a scalar,
    \bea
    \frac{i}{(p_i + q)^2 - m_j^2}
     =     \frac{i}{m_i^2 - m_j^2 + 2 p_i \cdot  q + \CO(q^2)},
           \label{propa}
    \eea
  where $m_j$ is the mass of the intermediate state that propagates. 
  A similar expression holds for a fermion.
  Thus, only when the masses are degenerate, $m_i = m_j$, (\ref{propa}) is singular as $q^{\mu} \rightarrow 0$.
  Under the assumption that there is no such degeneracy, Fig.~\ref{fig:NGF1}-2 does not contribute.
  The conservation of the current $\partial_\mu \langle B | \CS^\mu | A \rangle = 0$ leads to
    \bea
    \lim_{q^{\mu} \rightarrow 0} M_{1}
     =     0.
    \eea
  The processes of this type with the NGF emission are suppressed.
  
\begin{figure}
     \begin{center}
     \includegraphics[scale=0.6]{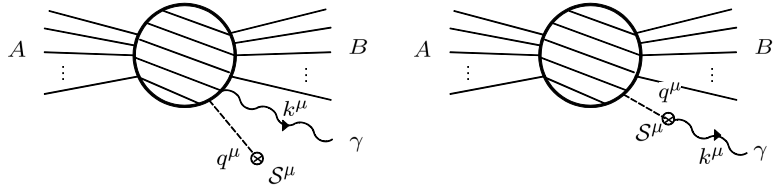}
     \caption{{\small Fig.~\ref{fig:NGF2}-1) 
                      $\CS^\mu$ is inserted in the end point of the NGF propagator (one-point coupling.)
                      Fig.~\ref{fig:NGF2}-2) 
                      $\CS^\mu$ is inserted in the external photon line (two-point coupling.)}}
     \label{fig:NGF2}
     \end{center}
     \end{figure}

  (2) Next, consider the processes of the second type with both photon emission and NGF emission.
  The diagrams with current insertion that give rise to singular contributions are illustrated in Fig.~\ref{fig:NGF2}.
  Contrary to the processes considered in (1), Fig.~\ref{fig:NGF2}-2 gives a nonvanishing contribution
  as both photon and the NGF are massless.
  The supercurrent (\ref{supercurrentN=1}) in fact contains a $\gamma$-$\lambda$ coupling,
  which is the second term in (\ref{supercurrentN=1}).
  Note that we can relate the radiative amplitude of this type with the amplitude $M_{1}$ discussed
  in (1) by the current conservation. 
  In fact, $\partial_\mu \langle B, \gamma |\CS^\mu| A \rangle$, leads to
    \bea
    \lim_{q \rightarrow 0} \epsilon^*_\mu M_{2}^\mu (q, k)
     =   - i f^{- 1}_{\lambda} \epsilon^*_\mu k_\nu \gamma^\mu \gamma^\nu \gamma_5 M_{1}(k).
    \eea

\end{document}